\documentclass[
aps,%
12pt,%
final,%
notitlepage,%
oneside,%
onecolumn,%
nobibnotes,%
nofootinbib,%
superscriptaddress,%
noshowpacs,%
centertags,%
noshowkeys,%
amssymb] {revtex4}
\usepackage{amsmath,graphicx}

\def\,{\ifmmode\mskip\thinmuskip\else\leavevmode\thinspace\fi}

\def\phi{\varphi}
\def\vare{\varepsilon}
\newcommand{\vecc}[1]{\mbox{\boldmath $#1$}}
\newcommand\nn{\nonumber}
\newcommand\dd{\mathrm{d}}
\newcommand\Tr{\mathrm{Tr\,}}
\newcommand\m[1]{\mathrm{#1}}

\begin{document}

\title{Calibration processes for photon-photon colliders}

\author{E.~Barto\v{s}}
\affiliation{Joint Institute for Nuclear Research, 141980 Dubna,
Russia}
\affiliation{Department of Theoretical Physics, Comenius
University, 84248 Bratislava, Slovakia.}
\author{A.-Z.~Dubni\v{c}kov\'a}
\affiliation{Department of Theoretical Physics, Comenius
University, 84248 Bratislava, Slovakia.}
\author{M.~V.~Galynskii}
\affiliation{Stepanov Institute of Physics BAS, 220072 Minsk,
Skorina ave. 70, Belarus}
\author{E.~A.~Kuraev}
\affiliation{Joint Institute for Nuclear Research, 141980 Dubna,
Russia}

\date{August 4, 2003}

\begin{abstract}
Processes with creation of a pair charged particles with emission
of hard photon and two pairs of charged particles are considered
for colliding partially polarized photon photon beams. The effects
of circular and linear polarization of the initial photons are
discussed in more details.
\end{abstract}

\maketitle

The planned $\gamma\gamma$ colliding beams based on laser backward
Compton scattering lepton high energy colliders \cite{Telnov} will
provide a new laboratory for investigation of   hadron properties.
In \cite{Ginzb} the general theory of polarization phenomena in
colliding photon beams was developed. So a lot of attention was
paid to the details of conversion of laser photons in the process
of backward Compton scattering and to the effects of density
distribution in the photon beams.

However, for the purposes of calibration and a measurement of the
degree of polarization of photon beams  the following  QED
processes with creation of one and two different pairs of leptons
\begin{subequations}
\begin{gather}
\gamma(k_1)+\gamma(k_2)\to \mu^+(q_+)+\mu^-(q_-), \label{eq:1a} \\
\gamma(k_1)+\gamma(k_2)\to \mu^+(q_+)+\mu^-(q_-)+\gamma(k),
\label{eq:1b}\\
\gamma(k_1)+\gamma(k_2)\to \bar{a}(p_+)+a(p_-)+\bar{b}(q_+)+b(q_-)
\label{eq:1c}
\end{gather}
\end{subequations}
can be utilized. Besides the latter, these processes provide an
essential background for study of  the hadron creation processes
as well as ones with heavy vector bosons. It has to be stressed,
that the polarization phenomena turns out to be essential in this
analysis.

A lot of attention \cite{Telnov} was paid to the processes
(\ref{eq:1a}, \ref{eq:1b}) as well as to the process (\ref{eq:1c})
\cite{KSS,KSSS}. In the one last especially the kinematics of the
main contribution to the total cross section was discussed more
carefully, namely when the final particles move in the narrow cones
along the direction of the photon colliding beams $\theta_i\sim
(M_i/\sqrt{s})$, $s=2k_1k_2$. Its total cross section do not decrease
as a function of the total cms energy $\sqrt{s}$.

\begin{figure} \label{fig:1}
\begin{center}
\includegraphics[scale=.65]{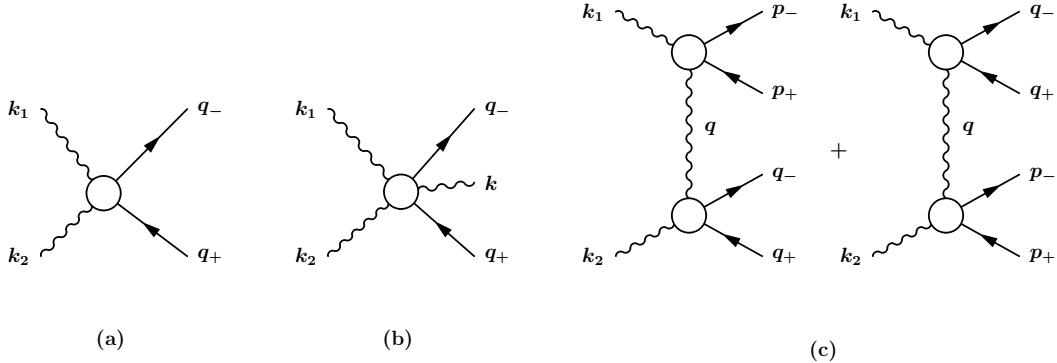}
\caption{Feynman diagram for the process
$\gamma(k_1)+\gamma(k_2)\to \mu^+(q_+)+\mu^-(q_-)$ (a), the
process $\gamma(k_1)+\gamma(k_2)\to
\mu^+(q_+)+\mu^-(q_-)+\gamma(k)$ (b) and the process
$\gamma(k_1)+\gamma(k_2)\to
e^+(p_+)+e^-(p_-)+\mu^+(q_+)+\mu^-(q_-) \,\,(c).$}
\end{center}
\end{figure}

The single pair creation processes (\ref{eq:1a}, \ref{eq:1b}) are
considered in the kinematical region, when the hard final
particles move in the large angles in cms
($\frac{m}{\sqrt{s}}\ll\theta_i\ll 1$), so the scalar products of
all different 4-vectors are large compared with the square masses
of particles. The process (\ref{eq:1c}) is investigated in the
quasi peripherical kinematical region, where the invariant masses
of  produced pairs are much smaller than $\sqrt{s}$ and large
compared with masses of particles. In our opinion this kinematics
is useful in the experiments because of absence of phone
processes. The rough estimation of differential cross sections is
$$\frac{\alpha^2}{s}\,,\;\frac{\alpha^3}{s}\,,\;
\frac{\alpha^4}{s_{max}}\,,$$ with $s_{max}=\max\{s_1,s_2\}$ and
$s_{1,\,2}$ are the  invariant masses squared of electron and muon
pairs respectively. So in a kinematical situation, when
$s_{max}\ll s$, the process (\ref{eq:1c}) is dominant.

The analysis of polarization phenomena in these all kinematical
regions (which is absent to our knowledge up to now) is the
motivation of our paper. Our paper is organized as follows: in
part 1 we consider the processes (\ref{eq:1a}, \ref{eq:1b}), in
part 2 the process (\ref{eq:1c}). The polarization effects are
studied by using  definite expressions for the chiral amplitudes.
We consider only the  kinematics (mentioned above) the invariant
mass of each pair is larger than  masses of particles and smaller
than cms total energy,  moving preferably along the one of initial
photons.

\vspace*{12pt} \noindent{\bf Process
$\gamma\gamma\rightarrow\mu^+\mu^-$ and
$\gamma\gamma\rightarrow\mu^+\mu^-\gamma$.}

The matrix elements of processes (\ref{eq:1a}, \ref{eq:1b}) are
(we put the masses of leptons to be equal to zero)
\begin{eqnarray}
M_{\lambda_1\lambda_2}^{\delta}=-i4\pi\alpha
\bar{u}_\delta(q_-)\left[\hat{\vare}_{\lambda_1}
\frac{\hat{q}_--\hat{k}_1} {-\chi_{1-}}\hat{\vare}_{\lambda_2}
+\hat{\vare}_{\lambda_2}\frac{-\hat{q}_+ +\hat{k}_1}{-\chi_{1+}}
\hat{\vare}_{\lambda_1}\right]v_\delta(q_+)\,
\end{eqnarray}
and
\begin{align} \label{eq:p2}
M_{\lambda_1\lambda_2}^{\lambda\delta}=&i(4\pi\alpha)^{3/2}
\bar{u}_{\delta}(q_-)\nn\\ \times &
\left[\hat{\vare}_{\lambda}^{\,*}\frac{\hat{q}_-+\hat{k}}{\chi_-}
\hat{\vare}_{\lambda_1}\frac{-\hat{q}_++\hat{k}_2}{-\chi_{2+}}
\hat{\vare}_{\lambda_2}+  %
\hat{\vare}_{\lambda_1}\frac{\hat{q}_--\hat{k}_1}{-\chi_{1-}}
\hat{\vare}_{\lambda}^{\,*}\frac{-\hat{q}_++\hat{k}_2}{-\chi_{2+}}
\hat{\vare}_{\lambda_2}\right. \nn \\
&+\hat{\vare}_{\lambda_1}\frac{\hat{q}_--\hat{k}_1}{-\chi_{1-}}
\hat{\vare}_{\lambda_2} \frac{-\hat{q}_+-\hat{k}}{\chi_+}
\hat{\vare}_{\lambda}^{\,*}+  %
\hat{\vare}_{\lambda}^{\,*}\frac{\hat{q}_-+\hat{k}}{\chi_-}
\hat{\vare}_{\lambda_2}\frac{-\hat{q}_+
+\hat{k}_1}{-\chi_{1+}}\hat{\vare}_{\lambda_1}  \\
&+\left.\hat{\vare}_{\lambda_2}\frac{\hat{q}_--\hat{k}_2}
{-\chi_{2-}}\hat{\vare}_{\lambda}^{\,*}\frac{-\hat{q}_++\hat{k}_1}
{-\chi_{1+}}\hat{\vare}_{\lambda_1}+ \hat{\vare}_{\lambda_2}
\frac{\hat{q}_--\hat{k}_2}{-\chi_{2-}}\hat{\vare}_{\lambda_1}
\frac{-\hat{q}_+-\hat{k}}{\chi_+}\hat{\vare}_{\lambda}^{\,*}\right]
v_{\delta}(q_+), \nn
\end{align}
with
\begin{equation}\label{}
\chi_{1\pm}=2k_1q_\pm\,,\quad \chi_{2\pm}=2k_2q_\pm\,,\quad
\chi_\pm=2kq_\pm\,\quad \lambda,\lambda_1,\lambda_2,\delta=\pm1,
\end{equation}
and $\hat{\vare}_{\lambda},\, \bar{u}_{\delta},\,v_\delta$ describe
the definite chiral states of photons
\begin{gather}
\hat{\vare}_\lambda(k)=N[\hat{q}_-\hat{q}_+\hat{k}\omega_{-\lambda}-
\hat{k}\hat{q}_-\hat{q}_+\omega_\lambda]\,,\quad
N=[s_1\chi_-\chi_+/2]^{-1/2}\,, \nonumber \\
\hat{\vare}_{\lambda_1}(k_1)=N_1[\hat{q}_-\hat{q}_+\hat{k}_1
\omega_{-\lambda_1}-
\hat{k}_1\hat{q}_-\hat{q}_+\omega_{\lambda_1}]\,,\quad
N_1=[s_1 \chi_{1-}\chi_{1+}/2]^{-1/2}\,,\\
\hat{\vare}_{\lambda_2}(k_2)=N_2[\hat{q}_-\hat{q}_+\hat{k}_2
\omega_{-\lambda_2}-\hat{k}_2\hat{q}_-\hat{q}_+\omega_{\lambda_2}]\,,\;
N_2=[s_1 \chi_{2-}\chi_{2+}/2]^{-1/2}\,,\nn \\
s_1=2q_-q_+\,,\quad\omega_{\pm \lambda}=(1\pm \lambda
\gamma_5)/2\, \nn
\end{gather}
and fermions \cite{Berends,GS}
\begin{eqnarray}
u_\delta=\omega_\delta u,\quad v_\delta=\omega_{-\delta}
v,\quad\bar{u}_\delta=\bar{u}\omega_{-\delta},\quad
\bar{v}_\delta=\bar{v}\omega_\delta\,.
\end{eqnarray}
For the process (\ref{eq:1a}) the chiral amplitudes
$M_{\lambda_1\lambda_2}^{\delta}$ are (we omit
$i(\sqrt{4\pi\alpha})^n$ factor)
\begin{gather}\label{eq:a1}
M_{+-}^\pm=\mp N_1N_2s\chi_{2\mp} \bar{u}(q_-)\hat{k}_1 \omega_\mp
v(q_+),\; \\ \nn %
M_{-+}^\pm=\pm N_1N_2s\chi_{2\pm}\bar{u}(q_-)\hat{k}_1\omega_\mp
v(q_+)\,,
\end{gather}
chiral amplitudes with equal chiralities of photons are equal
zero.

For a calculating of the cross sections for processes
(\ref{eq:1a}-\ref{eq:1c}) in the case of partially polarized
photon beams with momentum $k_i\, (i=1, 2)$ we will use
polarization matrices of density $\rho_i=\rho_i (k_i)$ in the
helically representation determined by Stokes parameters
$\vec{\xi}^{(i)}$  in the following way \cite{RQT}
\begin{equation}\label{eq:stox}
\rho_i=\rho_i (k_i) \; = \;\frac{1}{2}
\begin{pmatrix}
1+\xi_2^{(i)} & i \,\xi_1^{(i)}-\xi_3^{(i)}
\\-i\,\xi_1^{(i)}-\xi_3^{(i)} & 1-\xi_2^{(i)}
\end{pmatrix},\, \Tr(\rho_i)=1.
\end{equation}

\noindent Let us introduce $2\times2$ matrix building from the
amplitudes (\ref{eq:a1})
\begin{equation}\label{eq:m1}
{\cal M}_1^\delta =
\begin{pmatrix}
M_{++}^\delta & M_{+-}^\delta \\ M_{-+}^\delta & M_{--}^\delta
\end{pmatrix}.
\end{equation}
Than the probability of the process (\ref{eq:1a}) will be reduce
to calculation of trace from the product of the next matrices
\cite{Ginzb}
\begin{equation}\label{eq:tr1}
|M_{\lambda_1\lambda_2}^{\delta}|^2 \to \Tr(\rho_1^{T}\, {\cal
M}_1^\delta \,\rho_2 \,{{\cal M}_1^\delta}^\dagger)\,,
\end{equation}
where ${\rho_1^{T}}$ is the matrix transposed to $\rho_1$ and
${\cal M}_1^\dagger$ is hermitian conjugated matrix to ${\cal
M}_1$.

The cross section in general case has a form
\begin{align}
\frac{\dd\sigma^{\gamma\gamma\to\mu\bar{\mu}}_{
\vec{\xi}_1\vec{\xi}_2\delta}} {\dd\Omega_{\mu_-}}
=\frac{\alpha^2}{4s} \Big\{ (1 -\xi_2^{(1)} \xi_2^{(2)})R_+ -\,2\,
(\xi_1^{(1)} \xi_1^{(2)} +\xi_3^{(1)}
\xi_3^{(2)})+\delta(\xi_2^{(2)}-\xi_2^{(1)})R_- \Big\},
\label{stox1}\end{align} where
\begin{equation}
R_\pm=\frac{\chi_{1+}^2 \pm\chi_{1-}^2}{\chi_{1-}\chi_{1+}}\; , \;
\chi_{1\pm}=\chi_{2\mp}\,.
\end{equation}

For the case of completely circularly polarized photons (right
(R): $\xi_2^{(1,2)}=+1$, left (L): $\xi_2^{(1,2)}=-1$) we will
have
\begin{gather}
\frac{\dd\sigma^{\gamma\gamma\to\mu\bar{\mu}}_{LL}}{\dd\Omega_{\mu_-}}=
\frac{\dd\sigma^{\gamma\gamma\to\mu\bar{\mu}}_{RR}}{\dd\Omega_{\mu_-}}=0
\,,\quad
\frac{\dd\sigma^{\gamma\gamma\to\mu\bar{\mu}}_{LR}}{\dd\Omega_{\mu_-}}=
\frac{\dd\sigma^{\gamma\gamma\to\mu\bar{\mu}}_{RL}}{\dd\Omega_{\mu_-}}=
\frac{\alpha^2}{s} R_+\,.\nn
\end{gather}
Further when the colliding beams have equal or
reverse complete linear polarization $\xi^{(1)}_1=\pm
\xi^{(2)}_1=\pm1$ ($\xi^{(1)}_3=\pm \xi^{(2)}_3=\pm1$) from
(\ref{stox1}) we have
\begin{eqnarray}
\frac{\dd\sigma^{\gamma\gamma\to\mu\bar{\mu}}_{\pm
\pm}}{\dd\Omega_{\mu_-}} =\frac{\alpha^2}{2s} \left( R_+ \mp 2
\right)\, .\label{linpol} \nn
\end{eqnarray}
Finally in the case of unpolarized particles we have the result
which coincides with  \cite{RQT} in massless limit
\begin{eqnarray}
\frac{\dd\sigma^{\gamma\gamma\to\mu\bar{\mu}}}{\dd\Omega_{\mu_-}}
=\frac{\alpha^2}{2s}R_+\,.\nn
\end{eqnarray}

For the process (\ref{eq:1b}) chiral amplitudes (\ref{eq:p2})
could be written in the form
\begin{align}
&M_{++}^{-+}=NN_1N_2s_1^2\chi_+\bar{u}\hat{k}\omega_+v,\quad&
&M_{--}^{+-}=NN_1N_2s_1^2\chi_+\bar{u}\hat{k}\omega_-v,\nonumber\\
&M_{+-}^{++}=NN_1N_2s_1^2\chi_{2+}\bar{u}\hat{k}_2\omega_+v,\quad&
&M_{-+}^{--}=NN_1N_2s_1^2\chi_{2+}\bar{u}\hat{k}_2\omega_-v,\nonumber\\
&M_{-+}^{++}=NN_1N_2s_1^2\chi_{1+}\bar{u}\hat{k}_1\omega_+v,\quad&
&M_{+-}^{--}=NN_1N_2s_1^2\chi_{1+}\bar{u}\hat{k}_1\omega_-v,\\\nonumber
&M_{-+}^{-+}=-NN_1N_2s_1^2\chi_{2-}\bar{u}\hat{k}_2\omega_+v,\quad&
&M_{+-}^{+-}=-NN_1N_2s_1^2\chi_{2-}\bar{u}\hat{k}_2\omega_-v,\\\nonumber
&M_{+-}^{-+}=-NN_1N_2s_1^2\chi_{1-}\bar{u}\hat{k}_1\omega_+v,\quad&
&M_{-+}^{+-}=-NN_1N_2s_1^2\chi_{1-}\bar{u}\hat{k}_1\omega_-v,\\\nonumber
&M_{--}^{++}=-NN_1N_2s_1^2\chi_-\bar{u}\hat{k}\omega_+v,\quad&
&M_{++}^{--}=-NN_1N_2s_1^2\chi_-\bar{u}\hat{k}\omega_-v.
\end{align}

The matrix element squared for the process (\ref{eq:1b}) in the
case of partially polarized initial beams and summed over the
polarizations of final particles is calculated analogously to the
process (\ref{eq:1a}) (see (\ref{eq:tr1})). As a result for
differential cross section with taken into account only
polarization of photon beams  we have
\begin{gather} \label{eq:14}
\dd\sigma^{\gamma\gamma\to\mu \bar{\mu}\gamma}_{\vec{\xi}_1
\vec{\xi}_2}= \frac{\alpha^3s_1}{2\pi^2 s}\,
\frac{T_{in}}{D}\,\dd\Gamma\,,\quad
D=\chi_-\chi_+\chi_{1-}\chi_{1+}\chi_{2-}\chi_{2+}\,,\\
\dd\Gamma = \frac{\dd^3q_+}{\epsilon_+}\frac{\dd^3q_-}{\epsilon_-}
\frac{\dd^3k}{\omega}\delta^4(k_1+k_2-q_+-q_--k)\;, \nn
\end{gather}
\begin{align}
T_{in}&=\big(\xi_1^{(1)} \xi_1^{(2)}+ \xi_3^{(1)}
\xi_3^{(2)}\big)(\chi_{1+}\chi_{2+}+\chi_{1-}\chi_{2-})
(\chi_+\chi_- -\chi_{1+}\chi_{1-}-\chi_{2+}\chi_{2-})\nn\\
&+4\big(\xi_1^{(1)}\xi_3^{(2)}-\xi_3^{(1)} \xi_1^{(2)}\big)
(\chi_{1+}\chi_{2+}-\chi_{1-}\chi_{2-})\:E_q \nn\\
&- 4\xi_1^{(1)}(\chi_+ \chi_{2-}-\chi_- \chi_{2+})\:E_q
+4\xi_1^{(2)}(\chi_+ \chi_{1-}-\chi_- \chi_{1+})\:E_q \nn\\
&+\xi_3^{(1)}(\chi_+ \chi_{2-}+\chi_- \chi_{2+})
(\chi_+ \chi_- -\chi_{1+}\chi_{1-}+\chi_{2+}\chi_{2-})\nonumber\\
&+\xi_3^{(2)}(\chi_+ \chi_{1-}+\chi_- \chi_{1+})
(\chi_+\chi_- +\chi_{1+}\chi_{1-}-\chi_{2+}\chi_{2-})\nonumber \\
&+\xi_2^{(1)}\xi_2^{(2)}\big[\chi_+\chi_-(\chi_+^2+\chi_-^2)-
\chi_{1+}\chi_{1-}(\chi_{1+}^2+\chi_{1-}^2)
-\chi_{2+}\chi_{2-}(\chi_{2+}^2+\chi_{2-}^2)\big]\nn\\
&+\chi_+\chi_-(\chi_+^2+\chi_-^2)
+\chi_{1+}\chi_{1-}(\chi_{1+}^2+\chi_{1+}^2)+
\chi_{2+}\chi_{2-}(\chi_{2+}^2+\chi_{2-}^2)\;, \nn
\end{align}
where $E_q=\epsilon_{\mu\nu\rho\sigma}k_1^\mu k_2^\nu q_+^\rho
q_-^\sigma.=\frac{s}{2}[\vecc{q}_+\vecc{q}_-]_z$.

The corresponding cross sections for definite chiral states of
initial photons are

\begin{align}
\frac{\dd\sigma^{\gamma\gamma\to\mu^-\mu^+\gamma}_{RR(LL)}}
{\dd\Gamma}&=
\frac{\alpha^3s_1}{\pi^2s}\;
\frac{\chi_-\chi_+(\chi_-^2+\chi_+^2)}{D}\; ,
\\ \nonumber
\frac{\dd\sigma^{\gamma\gamma\to\mu^-\mu^+\gamma}_{RL(LR)}}
{\dd\Gamma}&= \frac{\alpha^3s_1}{\pi^2s}\;
\frac{\chi_{1-}\chi_{1+}(\chi_{1-}^2+\chi_{1+}^2)+\chi_{2-}\chi_{2+}
(\chi_{2-}^2+\chi_{2+}^2)}{D} \,.\nonumber
\end{align}%

For the case when only one photon beam polarized we gained the
cross section similar to (\ref{eq:14}) with the replacement
$T_{in}\to T_{in}^{(1)}$, where

\begin{align}
T^{(1)}_{in}=& \chi_+\chi_-(\chi_+^2+\chi_-^2)
+\chi_{1+}\chi_{1-}(\chi_{1+}^2+\chi_{1+}^2)\\ &+
\chi_{2+}\chi_{2-}(\chi_{2+}^2+\chi_{2-}^2)-4\xi_1^{(1)}(\chi_+
\chi_{2-}-\chi_- \chi_{2+})\:E_q\nn \\&+ \xi_3^{(1)}(\chi_+
\chi_{2-}+\chi_- \chi_{2+}) (\chi_+ \chi_-
-\chi_{1+}\chi_{1-}+\chi_{2+}\chi_{2-})\;. \nn
\end{align}
In the case when all particles are unpolarized the differential
cross section is
\begin{align}
\frac{\dd\sigma^{\gamma\gamma\to\mu^-\mu^+\gamma}_0}{\dd\Gamma}=&
\frac{\alpha^3s_1}{2\pi^2s}
\Big[\chi_-\chi_+(\chi_-^2+\chi_+^2)\\
&+\chi_{1-}\chi_{1+}(\chi_{1-}^2+\chi_{1+}^2)+\chi_{2-}\chi_{2+}
(\chi_{2-}^2+\chi_{2+}^2)\Big]/D\;.\nn
\end{align}

The probability of the process (\ref{eq:1b}) in the case when
initial and emitted photons with momenta $k_1$ and $k$ are
partially polarized with Stokes parameters $\vec{\xi}^{(1)}$ and
$\vec{\xi}$ is calculated analogously to (\ref{eq:tr1}).

Summing over the polarizations of final $\mu^+$ and $\mu^-$
particles one can get the differential cross section which takes
into account polarization of photons with momentum $k_1$ and $k$
\begin{equation}
\dd\sigma^{\gamma\gamma\to\mu\bar{\mu}\gamma}_{\vec{\xi}_1\vec{\xi}}
=\frac{\alpha^3s_1}{4\pi^2 s}\, \frac{T_{fin}}{D}\,\dd\Gamma
\label{stox2a} \;,
\end{equation}
\begin{align}
T_{fin}&=\big(\xi_1^{(1)} \xi_1+ \xi_3^{(1)} \xi_3\big)
(\chi_{+}\chi_{1+}+\chi_{-}\chi_{1-})
(\chi_{+}\chi_{-}+\chi_{1+}\chi_{1-}-\chi_{2+}\chi_{2-})\nonumber\\
&+4(\xi_1^{(1)}\xi_3-\xi_3^{(1)} \xi_1 )
(\chi_{+}\chi_{1+}-\chi_{-}\chi_{1-})\;E_q\nonumber \\
&-4 \xi_1^{(1)}(\chi_{+}\chi_{2-}-\chi_{-}\chi_{2+})\;E_q
-4\xi_1(\chi_{1+}\chi_{2-}-\chi_{1-}\chi_{2+})\;E_q \label{Tfin} \\
&+\xi_3^{(1)}(\chi_{+}\chi_{2-}+\chi_{-}\chi_{2+})
(\chi_{+}\chi_{-}-\chi_{1+}\chi_{1-}+\chi_{2+}\chi_{2-})\nonumber \\
&+\xi_3(\chi_{1+}\chi_{2-}+\chi_{1-}\chi_{2+})
(\chi_{+}\chi_{-}-\chi_{1+}\chi_{1-}-\chi_{2+}\chi_{2-})\nonumber \\
&+\xi_2^{(1)}\xi_2\big[\chi_{+}\chi_{-}(\chi_{+}^2+\chi_{-}^2)+
\chi_{1+}\chi_{1-}(\chi_{1+}^2+\chi_{1-}^2)
-\chi_{2+}\chi_{2-}(\chi_{2+}^2+\chi_{2-}^2)\big]\nonumber\\
&+\chi_{+}\chi_{-}(\chi_{+}^2+\chi_{-}^2)+%
\chi_{1+}\chi_{1-}(\chi_{1+}^2+\chi_{1-}^2)+
\chi_{2+}\chi_{2-}(\chi_{2+}^2+\chi_{2-}^2)\;. \nonumber
\end{align}%
Obtained expressions (\ref{stox2a}, \ref{Tfin}) allow us to
determine Stokes parameters of emitted photon versus polarization
of initial photon $\vec{\xi}^{(1)}$ with momentum $k_1$

\begin{gather}
\xi_1^f= \frac{f_{10}+\xi_1^{(1)}f_{11}+\xi_3^{(1)} f_{13}
}{f_{00}+\xi_1^{(1)} f_{01} +\xi_3^{(1)} f_{03}}\;,\quad\xi_3^f=
\frac{f_{30}+\xi_1^{(1)} f_{31}+\xi_3^{(1)} f_{33}
}{f_{00}+\xi_1^{(1)} f_{01}
+\xi_3^{(1)} f_{03}}\;, \\
\xi_2^f= \frac{\xi_2^{(1)} f_{22}} {f_{00}+\xi_1^{(1)} f_{01}
+\xi_3^{(1)} f_{03}}\;,\; \nonumber
\end{gather}
where
\begin{align}
f_{00}&=\chi_{+}\chi_{-}(\chi_{+}^2+\chi_{-}^2)+\chi_{1+}\chi_{1-}
(\chi_{1+}^2+\chi_{1-}^2)+\chi_{2+}\chi_{2-}(\chi_{2+}^2
+\chi_{2-}^2),\nonumber\\ %
f_{01}&=- 4(\chi_{+}\chi_{2-}-\chi_{-}\chi_{2+})\;E_q, \nonumber \\
f_{03}&=(\chi_{+}\chi_{2-}+\chi_{-}\chi_{2+})
(\chi_{+}\chi_{-}-\chi_{1+}\chi_{1-}+\chi_{2+}\chi_{2-}),\nonumber \\
f_{10}&=-4(\chi_{1+}\chi_{2-}-\chi_{1-}\chi_{2+})\;E_q, \nonumber \\
f_{11}&=(\chi_{+}\chi_{1+}+\chi_{-}\chi_{1-})
(\chi_{+}\chi_{-}+\chi_{1+}\chi_{1-}-\chi_{2+}\chi_{2-}), \nonumber \\
f_{13}&=-4(\chi_{+}\chi_{1+}-\chi_{-}\chi_{1-})\;E_q, \nonumber\\
f_{22}&=\chi_{+}\chi_{-}(\chi_{+}^2+\chi_{-}^2)+
\chi_{1+}\chi_{1-}(\chi_{1+}^2+\chi_{1-}^2)
-\chi_{2+}\chi_{2-}(\chi_{2+}^2+\chi_{2-}^2), \\
f_{30}&=(\chi_{1+}\chi_{2-}+\chi_{1-}\chi_{2+})
(\chi_{+}\chi_{-}-\chi_{1+}\chi_{1-}-\chi_{2+}\chi_{2-}), \nonumber \\
f_{31}&=-f_{13}=4(\chi_{+}\chi_{1+}-\chi_{-}\chi_{1-})\;E_q,\nonumber \\
f_{33}&=f_{11}=(\chi_{+}\chi_{1+}+\chi_{-}\chi_{1-})
(\chi_{+}\chi_{-}+\chi_{1+}\chi_{1-}-\chi_{2+}\chi_{2-})\:.\nonumber
\end{align}

\vspace*{36pt} \noindent{\bf Process $\gamma\gamma\rightarrow
a(p_-)\bar{a}(p_+)b(q_-)\bar{b}(q_+)$.}

The kinematical variables of two pairs production process in quasi
peripherical photon collisions are defined as (see (\ref{eq:1c}))
\begin{gather}
2k_1 k_2=s,\quad (p_++p_-)^2=s_1\,,\quad (q_++q_-)^2=s_2\,,\\
\nonumber q=k_1-p_+-p_-=q_++q_--k_2\,,\quad
\chi_\pm=2k_1p_\pm\,,\quad \chi'_\pm=2k_2 q_\pm\,.
\end{gather}
Further we will consider the case when the $\m{e}^+\m{e}^-$ pair
moves in the same hemisphere with the photon of the  momentum $k_1$
and the muon pair moves with another photon in an opposite
hemisphere. Besides the latter we will restrict ourselves by the case
when the invariant masses $\m{e}^+\m{e}^-$ and $\mu^+\mu^-$ are large
in comparison with the mass of muon and much less than the center of
mass total energy $\sqrt{s}$
\begin{eqnarray}
m_\mu^2\ll s_1,\quad s_2\ll s,\quad \chi_\pm\sim s_1,\quad
\chi'_\pm\sim s_2.
\end{eqnarray}
A contribution of this region will not depend on $s$ at large $s$,
and will be dominant. For this kinematics the components of the
created pairs  move inside the cones with polar angle of order
$\theta_{1,\,2}\sim\sqrt{s_{1,\,2}/s}$ along the beam axes. A few
Feynman amplitudes are relevant in this kinematics, which can be
drawn in the form of two block  (see Fig.~1). The corresponding
matrix element has a factorized form
\begin{gather} \label{eq:matrix}
M(\gamma\gamma\rightarrow \m{e}^+\m{e}^-\mu^+\mu^-)=
is\frac{2(4\pi\alpha)^2}{q^2}
\big[m_1^{\lambda_1\lambda_e}m_2^{\lambda_2\lambda_\mu} +
m_1^{\lambda_2\lambda_e}m_2^{\lambda_1\lambda_\mu}\big],\\ \nn
m_1^{\lambda_1\lambda_e}=\frac{1}{s}k_2^\mu
\vare_1^\nu(k_1)M_{1\mu\nu},\quad
m_2^{\lambda_2\lambda_\mu}=\frac{1}{s}k_1^\sigma
\vare_2^\rho(k_2)M_{2\sigma\rho}.
\end{gather}
In this two back-to-back kinematical regions the Sudakov
parametrization for 4-momenta is convenient \cite{Sudakov}.
Assuming that the pair $a(p_-)\bar{a}(p_+)$ belongs to the jet
moving along direction of photon $k_1$ we have
\begin{gather}
p_\pm=\alpha_\pm k_2+x_\pm k_1+p_\pm^\bot,\quad
(p_\pm^\bot)^2=-\vecc{p}_\pm^2,\quad
p_\pm^\bot.k_1=p_\pm^\bot.k_2=0,
\\ \nn \alpha_\pm\approx\frac{\chi_{1\pm}}{s}=
\frac{\vecc{p}_\pm^2}{sx_\pm},\quad s_1=(p_++p_-)^2=
\frac{\vecc{Q_a}{\kern-.4em\hbox{}^{2}}} {x_+x_-},\quad
\vecc{Q_a}=x_+\vecc{p}_--x_-\vecc{p}_+,
\end{gather}
where $\vecc{p}_\pm$ are 2-dimensional euclidean vectors
$(\vecc{p}_-+\vecc{p}_++\vecc{q}=0)$, $x_\pm=2k_2p_\pm/s$ are
energy fractions of pair components ($x_++x_-=1$) and $s_1$ is
squared invariant mass of the pair.

The similar relations are valid for the pair
$b(q_-)\,\bar{b}(q_+)$ from the opposite jet moving in the
direction of the photon $k_2$
\begin{gather}
q_\pm=y_\pm k_2+\beta_\pm k_1+q_\pm^\bot\,,\quad
q_\pm^\bot=-\vecc{q}_\pm^2\,,\quad
q_\pm^\bot.k_1=q_\pm^\bot.k_2=0\,,
\\ \nn \beta_\pm\approx\frac{\chi_{2\pm}'}{s}=
\frac{\vecc{q}_\pm^2}{sy_\pm}\,,\quad s_2=(q_++q_-)^2=
\frac{\vecc{Q_b}{\kern-.2em\hbox{}^{2}}}{y_+y_-}\,,\quad
\vecc{Q_b}=y_+\vecc{q}_--y_-\vecc{q}_+\,,
\end{gather}
where energy fractions $y_\pm=2k_1q_\pm/s$, $(y_++y_-=1)$ and
$\vecc{q}_-+\vecc{q}_+-\vecc{q}=0$.

We define the chiral amplitudes as a matrix elements calculated
with definite chiral states of fermions and photons. We choice the
polarization vectors of photons in the form \cite{Berends}
\begin{equation}\label{}
\vare^{\lambda_j}_{\mu}(k)=\vare_\mu^{||}+i{\lambda_j} \vare_\mu^\bot
\end{equation}
\begin{gather}
\vare_\mu^{||}=N_2[(q_-k) q_{+\mu}-(q_+k) q_{-\mu}]\,,\quad
\vare_\mu^\bot=N_2\epsilon_{\mu\alpha\beta\gamma}
q_-^\alpha q_+^\beta k^\gamma\,, \nonumber \\
\hat{\vare}^{\lambda_1}=N_1[\hat{p}_-\hat{p}_+\hat{k}_1
\omega_{-{\lambda_1}} -\hat{k}_1\hat{p}_-\hat{p}_+
\omega_{\lambda_1}]\,,\quad
N_1=[s_1\chi_+\chi_-/2]^{-1/2}\,, \nn \\
\hat{\vare}^{\lambda_2}=N_2[\hat{q}_-\hat{q}_+\hat{k}_2
\omega_{-{\lambda_2}}- \hat{k}_2\hat{q}_-\hat{q}_+
\omega_{\lambda_2}]\,,\quad N_2=[s_2\chi'_+\chi'_-/2]^{-1/2}\,.
\nonumber
\end{gather}
Chiral states of fermions were defined above.

In derivation of cross section we consider only first term in
(\ref{eq:matrix}), the second one can be obtained by correspondent
replacement. Their interference term vanish in limit $s\to\infty$
and is desregarded.
\begin{eqnarray}
M^{\lambda_1\lambda_e\lambda_2\lambda_\mu}=
is\frac{2(4\pi\alpha)^2}{\vecc{q}^2}
m_1^{\lambda_1\lambda_e}m_2^{\lambda_2\lambda_\mu},\;
\lambda_j=\pm 1,\; j=1,2,e,\mu.
\end{eqnarray}
The quantities $m_1^{\lambda_1\lambda_e}$ for lepton pair have a
form
\begin{equation}\label{eq:m}
m_1^{\lambda_1\lambda_e}=\frac{N_1}{s}\bar{u}(p_-)\left[
\delta_{\lambda_e\lambda_1}\hat{p}_+\hat{q}\hat{k}_2+
\delta_{\lambda_e(-\lambda_1)}\hat{k}_2\hat{q}\hat{p}_-\right]
\omega_{\lambda_1} v(p_+),
\end{equation}
For the case of creation of charged pion pair $\pi^+\pi^-$ one
gets
\begin{equation}\label{}
m_1^\lambda=-N_2\big\{\vecc{Q_a}.\vecc{q}-
i\lambda[\vecc{Q_a},\vecc{q}]_z\big\}.
\end{equation}
Quantities $m_2^{\lambda_2\lambda_\mu}$ are constructed
analogically.

Let us introduce $2\times2$ matrices building from the amplitudes
(\ref{eq:m})
\begin{subequations}
\begin{align}
{\cal M}_1^{\lambda_e} &=
\begin{pmatrix}
m_{+}m_{+}^* & m_{+}m_{-}^* \\
m_{-}m_{+}^* & m_{-}m_{-}^*
\end{pmatrix},
\quad m_{\pm}=
m_1^{\lambda_1 \lambda_e}\;\;(\lambda_1=\pm 1 ), \\
{\cal M}_2^{\lambda_\mu} &=
\begin{pmatrix}
m'_{+}{m'_{+}}^* & m'_{+}{m'_{-}}^* \\
m'_{-}{m'_{+}}^* & m'_{-}{m'_{-}}^*
\end{pmatrix},
\quad m'_{\pm}= m_2^{\lambda_2 \lambda_\mu}\;\; (\lambda_2=\pm 1),
\end{align}
\end{subequations}
then the absolute values of matrix elements squared  of the
process (\ref{eq:1c}) will be reduce to calculation of a trace
from the product of the  matrices as follows
\begin{equation}\label{}
\big|m_1^{\lambda_1\lambda_e}\big|^2=\Tr(\rho_1\, {\cal
M}_1^{\lambda_e}), \quad
\big|m_2^{\lambda_2\lambda_\mu}\big|^2=\Tr(\rho_2 \,{\cal
M}_2^{\lambda_\mu})\,.
\end{equation}
A simple calculation leads to the result
\begin{subequations}
\begin{align}
&{\cal M}_{1(e)}^\pm
=N_1^2\vecc{Q_a}{\kern-.4em\hbox{}^{2}}\vecc{q}^2
\begin{pmatrix}
{x_\pm}/{x_\mp} & \exp\{\pm 2i\phi_a\} \\
\exp\{\mp 2i\phi_a\} & {x_\mp}/{x_\pm}
\end{pmatrix},\\
&{\cal M}_{1(\pi)}
=N_1^2\vecc{Q_a}{\kern-.4em\hbox{}^{2}}\vecc{q}^2
\begin{pmatrix}
1 & \exp\{2i\phi_a\} \\
\exp\{-2i\phi_a\} & 1
\end{pmatrix},\;\;\; \phi_a=\widehat{\vecc{Q_aq}}.
\end{align}
\end{subequations}

Let us now transform the final state phase space volume
\begin{align}\label{eq:delta}
\dd\Phi_4&=\frac{\dd^3p_+}{2p_{+0}}\:\frac{\dd^3p_-}{2p_{-0}}\:
\frac{\dd^3q_+}{2q_{+0}}\: \frac{\dd^3q_-}{2q_{-0}}\:
\delta^4(k_1+k_2-q_+-q_--p_+-p_-) \\
&= \dd^4q\:\dd^4p_+\:\dd^4p_-\:\dd^4q_+\:\dd^4q_-\:
\delta^4(k_1+q-p_+-p_-) \delta^4(k_2-q-q_+-q_-)\nn \\ \nn &\quad
\times\delta(p_+^2)\:\delta(p_-)\: \delta(q_+^2)\:\delta(q_-^2).
\end{align}
Applying Sudakov representation ($\dd^4q=\frac{s}{2}\dd\alpha
\dd\beta \dd^2\vecc{q}$) one can rewrite (\ref{eq:delta}) into the
following form
\begin{equation}\label{eq:phase}
\dd\Phi_4=\frac{\dd x_-\:\dd y_-\:\dd^2q_+\:\dd^2q_-\:\dd^2q}
{8s\:x_+\:x_-\:y_+\:y_-}.
\end{equation}
For the cross section of two different pairs production (i.e.,
$a\neq b$) one obtains
\begin{align}
\dd\sigma(\gamma\gamma\to a\bar{a}b\bar{b})=&
\frac{\alpha^4}{\pi}x_+x_-y_+y_- \frac{\dd x_-\dd
y_-\dd^2q_-\dd^2p_-\dd^2q}{\vecc{p}_-^2\vecc{q}_-^2} \\ \nn
&\times\bigg[\frac{S_{1(a)}S_{2(b)}}{(\vecc{q}+\vecc{q}_-)^2
(\vecc{q}-\vecc{p}_-)^2}+
\frac{S_{1(b)}S_{2(a)}}{(\vecc{q}-\vecc{q}_-)^2
(\vecc{q}+\vecc{p}_-)^2}\bigg],
\end{align}
and for the case $a=b$
\begin{align}
\dd\sigma(\gamma\gamma\to a\bar{a}a\bar{a})=&
\frac{\alpha^4}{\pi}x_+x_-y_+y_- \frac{\dd x_-\dd
y_-\dd^2q_-\dd^2p_-\dd^2q}{\vecc{p}_-^2\vecc{q}_-^2} \\ \nn
&\times\frac{S_{1(a)}S_{2(a)}}{(\vecc{q}+\vecc{q}_-)^2
(\vecc{q}-\vecc{p}_-)^2}
\end{align}
where
\begin{equation} \label{eq:esy}
S_{1(a)}=\frac{\Tr(\rho_1\, {\cal M}_{1(a)})}
{\vecc{q}^2\vecc{Q}^2N_1^2}\,, \quad %
S_{2(a)}=\frac{\Tr(\rho_2\, {\cal M}_{2(a)})}
{\vecc{q}^2\vecc{Q}^2N_2^2}\,.
\end{equation}

Explicit form of (\ref{eq:esy}) is
\begin{align}
&S_{1(\pi)}=1-\xi_3^{(1)}\cos 2\phi_a+\xi^{(1)}_1\sin 2\phi_a\,,\nn\\
&S_{2(\pi)}=1-\xi_3^{(2)}\cos 2\phi_b-\xi^{(2)}_1\sin 2\phi_b\,,\\
&S_{1(\mu)}=\frac{x_-^2+x_+^2}{2x_+x_-}-
\lambda_\mu\xi^{(1)}_2\frac{x_--x_+}{2x_+x_-}- \xi^{(1)}_3\cos
2\phi_a+
\xi^{(1)}_1\sin 2\phi_a\,, \nn\\
&S_{2(\mu)}=\frac{y_-^2+y_+^2}{2y_+y_-}-
\lambda_\mu\xi_2^{(2)}\frac{y_--y_+}{2y_+y_-}- \xi_3^{(2)}\cos
2\phi_b+
\xi_1^{(2)}\sin 2\phi_b\,, \nn
\end{align}
with angles $\phi_j=\widehat{\vecc{Q_jq}}$.

In conclusion we note that the cross section of two pair
production in kinematical region, when all hard particles move in
large angles even in unpolarized case, has a very complicated form
(see for example the paper \cite{KP}, where it was obtained for a
cross channel). The ratio of magnitudes of cross sections in
peripherical kinematics to the last one is $\frac{s}{s_{max}}\gg
1$, which underline the importance of quasi peripherical
kinematics, considered in this paper.

\begin{acknowledgments}
Two of us (E.B., E.K.) are grateful to Institute of Physics SAS
(Bratislava), where part of this work was done. We are also
grateful to grant RFBR No. 03--02--17077 and grant INTAS No.
00366. The work was in part supporrted by Slovak Grant Agency for
Sciences, grant No. 2/1111/23 (A.-Z.D) and the grant BRFFI No.
F03-183 (M.G.).
\end{acknowledgments}

\end{document}